# A LANGUAGE INDEPENDENT WEB DATA EXTRACTION USING VISION BASED PAGE SEGMENTATION ALGORITHM

## [1]P YesuRaju, [2]P KiranSree

[1]PG Student, [2]Professorr, Department of Computer Science, B.V.C.E.College, Odalarevu, Andhra Pradesh, India
yesuraju.p@gmail.com, profkiran@yahoo.com

## Abstract

*Web usage mining is a process of extracting useful information from server logs i.e. user's history. Web usage mining is a process of finding out what users are looking for on the internet. Some users might be looking at only textual data, where as some others might be interested in multimedia data. One would retrieve the data by copying it and pasting it to the relevant document. But this is tedious and time consuming as well as difficult when the data to be retrieved is plenty. Extracting structured data from a web page is challenging problem due to complicated structured pages. Earlier they were used web page programming language dependent; the main problem is to analyze the html source code. In earlier they were considered the scripts such as java scripts and cascade styles in the html files. When it makes different for existing solutions to infer the regularity of the structure of the WebPages only by analyzing the tag structures. To overcome this problem we are using a new algorithm called VIPS algorithm i.e. independent language. This approach primary utilizes the visual features on the webpage to implement web data extraction.*

-------------------------------------------------------------***-------------------------------------------------------------

## 1. INTRODUCTION

Information drives today's businesses and the Internet is a powerhouse of information. Most businesses rely on the web to gather data that is crucial to their decision making processes. Companies regularly assimilate and analyze product specifications, pricing information, market trends and regulatory information from various websites and when performed manually, this is often a time consuming, error-prone process.

Automation Anywhere can help you easily automate data extraction without any programming. Going beyond simple screen scraping or cutting and pasting information from a website, Automation Anywhere intelligently extracts information. Running on SMART Automation Technology®, it can automatically login to websites, account for changes in the source website, extract that information and copy it to another application reliably in a format specified by you.

## 2 .RELATED WORK

A number of approaches have been reported in the literature for extracting information from Web pages. We briefly review earlier works based on the degree of automation in Web data extraction, and compare our approach with fully automated solutions since our approach belongs to this category.

### Manual Approaches

Some of the best known tools that adopt manual approaches are Minerva, TSIMMIS, and Web-OQL [1]. Obviously, they have low efficiency and are not scalable.

### Automatic Approaches

In order to improve the efficiency and reduce manual efforts, most recent researches focus on automatic approaches instead of manual ones. Some representative automatic approaches are Omini [2], Roadrunner, IEPAD, MDR, DEPTA.

## 3. VIPS

VIPS (vision based page segmentation algorithm) is an automatic top-down, tag tree independent approach to detect web content structure. VIPS algorithm is to transform a deep web page into a visual block tree. A visual block tree is actually a segmentation of a web page. The root block represents the whole page, and each block in the tree corresponds to a rectangular region on the web pages. The leaf blocks are the blocks that cannot be segmented further, and they represent the minimum semantic units, such as continuous texts or images. These block tree is constructed by using DOM (document object model) tree. There is a one main building component in the VIPS algorithm that is DOM (document object model) tree. The DOM tree is used to manage XML data or access a complex data structure repeatedly. The DOM is used to Builds the data as a tree structure in memory, Parses an entire XML document at one time, Allows applications to make dynamic updates to the tree structure in memory. (As a result, you could use a second application to create a new XML document based on the updated tree structure that is held in memory).An XML document is a string of characters. Almost every legal Unicode character may appear in an XML document. The processor analyzes the markup and passes structured information to an application. The specification places requirements on what an XML processor must do and not do, but the application is outside its scope. The processor (as the specification calls it) is often referred to colloquially as an XML parser. The characters which make up an XML document are divided into markup and content. Markup and content may be distinguished by the application of simple syntactic rules. All strings which constitute markup either begin with the character "<" and end with a ">", or begin with the character "&" and end with a ";". Strings of characters which are not markup are content.HTML, which stands for HyperText Markup Language, is the predominant markup language for web pages. HTML is the basic building-blocks of webpages.HTML is written in the form of HTML elements consisting of tags, enclosed in angle brackets (like <html>), within the web page content. HTML tags normally come in pairs like <h1> and </h1>. The first tag in a pair is the start tag, the second tag is the end tag (they are also called opening tags and closing tags). In between these tags web designers can add text, tables, images, etc.The purpose of a web browser is to read HTML documents and compose them into visual or audible web pages. The browser does not display the HTML tags, but uses the tags to interpret the content of the page.HTML elements form the building blocks of all websites. HTML allows images and objects to be embedded and can be used to create interactive forms. It provides a means to create structure documents by denoting structural semantics for text such as headings, paragraphs, lists, links, quotes and other items. It can embed scripts in languages such as JavaScript which affect the behavior of HTML WebPages. Web usage mining is a process of extracting useful information from server logs i.e. users history. Web usage mining is the process of finding out what users are looking for on the Internet. Some users might be looking at only textual data, whereas some others might be interested in multimedia data. One would retrieve the data by copying it and pasting it to the relevant document. But this is tedious and time-consuming as well as difficult when the data to be retrieved is plenty. This is when Web Data Extraction comes into play. The world web has close to one million searchable information according to recent survey. This searchable information include both search engine web databases. If you give query to the search engine the useful information from them can be retrieved. Normally the WebPages have images, links and data. WebPages are designed by using html files and xml files. Now a days the web page designers are increasing the complexity of html source code. So we will use VIPS algorithm and we will extract the data easily.

## 4. DESIGN

In Earlier work depends primarily on the programming languages, the challenges lies in analyzing the HTML code. In this project we are going to discuss about the VIPS algorithm. By using this algorithm to transform a web page into a visual block tree. A visual block tree is actually segmentation of a webpage. This VIPS algorithm is an automatic top-down; tag tree independent approach to detect web content structure.Basically, the vision-based content structure is obtained by using DOM structure. In this algorithm we follow three steps first one is block extraction, separator detection and content structure construction. These three as a whole regarded as a round. The algorithm is top-down. The web page is firstly segmented into several big blocks and the hierarchical structure of this level is recorded. For each block, the segmentation process is carried out recursively until we get sufficient small blocks.

The visual information of web pages, which has been introduced above, can be obtained through the programming interface provided by web browsers. In this paper, we employs the VIPS algorithm to transform a deep web page into a visual block tree .A visual block tree is actually a segmentation of a web page. The root block represents the whole page, and each block in the tree corresponds to a rectangular region on the web pages. The leaf blocks are the blocks that cannot be segmented further, and they represent the minimum semantic units, such as continuous texts or images. These visual block tree is constructed by using DOM tree. DOM tree means document object model. Therefore these are all about the design part of the visual block tree and after that we will extract images, links and data.

## 5. IMPLEMENTATION

In this section we are going to implement the DOM tree in order to find out the visual block tree.

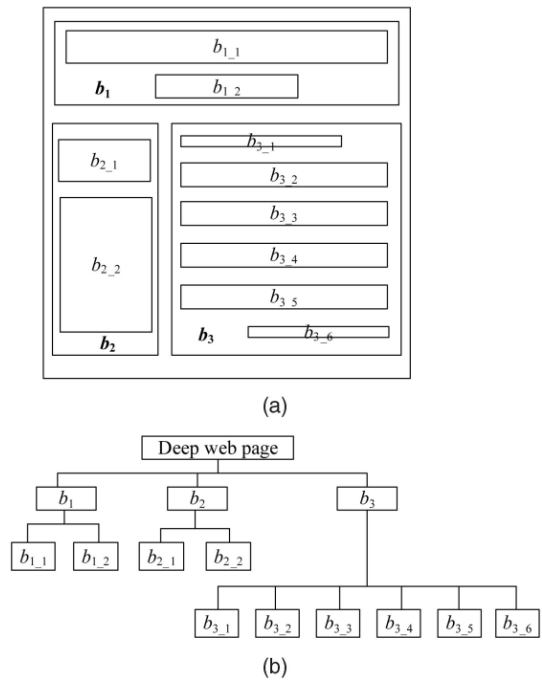

(a)

(b)

**Fig 1**.(a) The presentation structure, (b) its visual block tree.

## DOM TREE

In VIPS algorithm we will use DOM tress to find out the visual block tree. The Document Object Model (DOM) is a cross-platform and language-independent convention for representing and interacting with objects in HTML, XHTML and XML documents. Aspects of the DOM (such as its "Elements") may be addressed and manipulated within the syntax of the programming language in use. The public interface of a DOM is specified in its application programming interface (API).

The DOM is a programming API for documents. It is based on an object structure that closely resembles the structure of the documents it models. For instance, consider this table, taken from an HTML document. In this we will take a sample html code and converted into a DOM tree

```
 <TABLE>
<TBODY>
<TR>
<TD>Shady Grove</TD>
<TD>Aeolian</TD>
</TR>
<TR>
<TD>Over the   River, Charlie</TD>
<TD>Dorian</TD>
</TR>
</TBODY>
</TABLE>
```

A graphical representation of the DOM tree of the above html code is given as below.

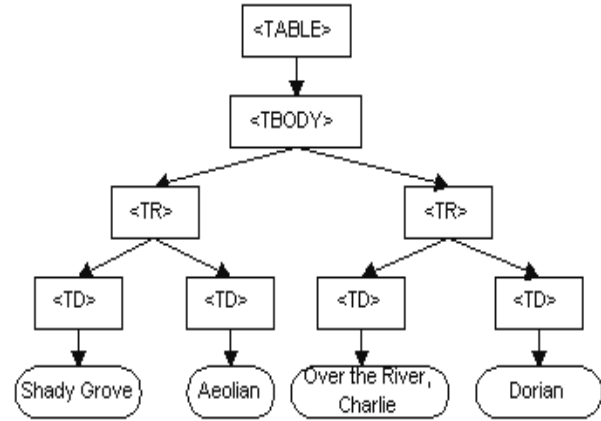

**Fig2:** Graphical representation of the DOM of the example table.

In the DOM, documents have a logical structure which is very much like a tree, to be more precise, which is like a "forest" or "grove", which can contain more than one tree. Each document contains zero or one doctype nodes, one root element node, and zero or more comments or processing instructions; the root element serves as the root of the element tree for the document. However, the DOM does not specify that documents must be implemented as a tree or a grove, nor does it specify how the relationships among objects be implemented. The DOM is a logical model that may be implemented in any convenient manner. In this specification, we use the term structure model to describe the tree-like representation of a document. We also use the term "tree" when referring to the arrangement of those information items which can be reached by using "tree-walking" methods; (this does not include attributes). One important property of DOM structure models is structural isomorphism. If any two Document Object Model implementations are used to create a representation of the same document, they will create the same structure model, in accordance with the XML Information Set.

## HTML DOM

The DOM defines a standard for accessing documents like HTML and XML.

The DOM is separated into 3 different parts  levels:

•     Core DOM - standard model for any structured document

•     XML DOM - standard model for XML documents

•     HTML DOM - standard model for HTML documents
The html dom is a standard object model for any structured document, a standard interface for programming interface html, platform and language independent. The dom says The entire document is a document node Every HTML element is an element node, The text in the HTML elements are text nodes, Every HTML attribute is an attribute node Comments are comment nodes. The HTML DOM views a HTML document as a tree-structure. The tree structure is called a node-tree. All nodes can be accessed through the tree. Their contents can be modified or deleted, and new elements can be created. The node tree below shows the set of nodes, and the connections between them. The tree starts at the root node and branches out to the text nodes at the lowest level of the tree. The HTML DOM views a HTML document as a node-tree. All the nodes in the tree have relationships to each other.

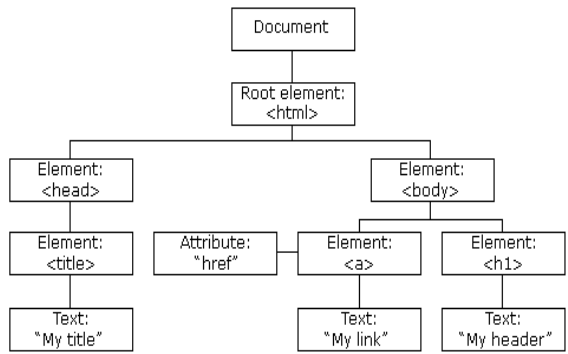

**Figure3:** Html Dom Node tree

The nodes in the node tree have a hierarchical relationship to each other. The terms parent, child, and sibling are used to describe the relationships. Parent nodes have children. Children on the same level are called siblings (brothers or sisters).

- In a node tree, the top node is called the root
- Every node, except the root, has exactly one parent node
- A node can have any number of children
- A leaf is a node with no children
- Siblings are nodes with the same parent

You can access a node in three ways By using the getElementById () method, By using the getElementsByTagName () method and By navigating the node tree, using the node relationships.

## XML DOM

The XML DOM is a standard object model for XML, a standard programming interface for XML, Platform- and language-independent. The XML DOM defines the objects and properties of all XML elements, and the methods (interface) to access them.

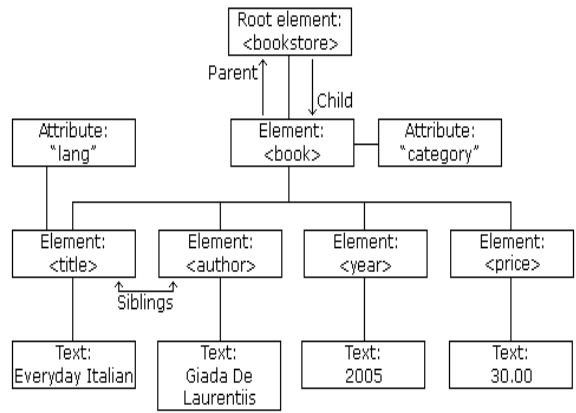

**Figure4:** Xml dom tree node

The XML DOM views an XML document as a tree-structure. The tree structure is called a node-tree. All nodes can be accessed through the tree. Their contents can be modified or deleted, and new elements can be created. The node tree shows the set of nodes, and the connections between them. The tree starts at the root node and branches out to the text nodes at the lowest level of the tree.

The XML DOM contains methods (functions) to traverse XML trees, access, insert, and delete nodes. However, before an XML document can be accessed and manipulated, it must be loaded into an XML DOM object. An XML parser reads XML, and converts it into an XML DOM object that can be accessed with JavaScript. Most browsers have a built-in XML parser. For security reasons, modern browsers do not allow access across domains. This means, that both the web page and the XML file it tries to load, must be located on the same server.

A web browser typically reads and renders HTML documents. This happens in two phases: the *parsing phase* and the *rendering phase*. During the parsing phase, the browser reads the markup in the document, breaks it down into components, and builds a document object model (DOM) tree. By using this VIPS algorithm we will separate the links, images and the data very easily and then we will extract that links, images and the data very easily.

## CONCLUSION

In this paper we have proposed the VIPS algorithm which helps us to extract the data easily from the web page. Earlier they had used web page programming language dependent that is very difficult to analyze the data because of complicated html and xml structures. So we will extract the data easily by using this VIPS algorithm.